\documentclass[aps,prd,twocolumn,showpacs,preprintnumbers,nofootinbib]{revtex4}
\usepackage{amssymb,latexsym}
\usepackage{amsmath,amsbsy}
\usepackage{epsfig,bm}
\usepackage{feynmf}
\DeclareMathOperator{\pperp}{\mathbf{p}^{\perp}}

\DeclareMathOperator{\Pplus}{P^{+}}
\DeclareMathOperator{\pplus}{p^{+}}
\DeclareMathOperator{\qperp}{\mathbf{q}^{\perp}}

\DeclareMathOperator{\kperp}{\mathbf{k}^{\perp}}
\DeclareMathOperator{\kplus}{k^{+}}

\DeclareMathOperator{\ie}{i \epsilon}

\DeclareMathOperator{\dw}{D_{\text{W}}}
\DeclareMathOperator{\pp}{p^{\prime}}
\DeclareMathOperator{\GDA}{\Phi(z, \zeta, s)}

\DeclareMathOperator{\kminus}{k^{--}}
\DeclareMathOperator{\pminus}{p^{--}}

\DeclareMathOperator{\qplus}{q^+}
\DeclareMathOperator{\Gt}{\tilde{G}}
\DeclareMathOperator{\pe}{\mathcal{P}}
\DeclareMathOperator{\qu}{\mathcal{Q}}
\DeclareMathOperator{\psiket}{|\psi\rangle}
\DeclareMathOperator{\psitwoket}{|\psi_{2}\rangle}
\DeclareMathOperator{\psiquket}{|\psi_{\qu}\rangle}

\DeclareMathOperator{\psitwobra}{\langle\psi_{2}|}

\DeclareMathOperator{\dperp}{\mathbf{\Delta}^\perp}
\DeclareMathOperator{\kpp}{\mathbf{k}^{\prime\perp}}
\DeclareMathOperator{\ppp}{\mathbf{p}^{\prime\perp}}
\DeclareMathOperator{\xp}{x^\prime}
\DeclareMathOperator{\yp}{y^\prime}
\DeclareMathOperator{\xpp}{x^{\prime\prime}}
\DeclareMathOperator{\kppp}{\mathbf{k}^{\prime\prime\perp}}
\DeclareMathOperator{\zp}{z^\prime}
\DeclareMathOperator{\zpp}{z^{\prime\prime}}
\DeclareMathOperator{\ypp}{y^{\prime\prime}}

\begin{document}
\preprint{NT-UW 02-31}
\title{Current in the light-front Bethe-Salpeter formalism II:\\ Applications}
\author{B.~C.~Tiburzi}
%\email{bctiburz@u.washington.edu}
\author{G.~A.~Miller}
%\email{miller@phys.washington.edu}
\affiliation{Department of Physics  
	University of Washington      
	Box 351560
	Seattle, WA 98195-1560}
\date{\today}

\begin{abstract}
We pursue applications of the light-front reduction of current matrix elements in the Bethe-Salpeter formalism.
The normalization of the reduced wave function is derived from the covariant framework and related to non-valence probabilities 
using familiar Fock space projection operators. Using a simple model, we obtain expressions for generalized parton 
distributions that are continuous. The non-vanishing of these distributions at the crossover between kinematic regimes (where the plus component 
of the struck quark's momentum is equal to the plus component of the momentum transfer) is tied to higher Fock components. 
Moreover continuity holds due to relations between Fock components at vanishing plus momentum.
Lastly we apply the light-front reduction to time-like form factors and derive expressions for the generalized distribution amplitudes in this model. 
\end{abstract}


\maketitle

\section{Introduction}
More than a half century ago, Dirac's paper on the forms of relativistic dynamics \cite{Dirac:1949cp} introduced the front-form
Hamiltonian approach.  Applications to quantum mechanics and field theory were overlooked at the time due to the 
appearance of covariant perturbation theory.
%, in which the special role played by time was supplanted by Lorentz invariance. 
%Front-form dynamics gradually resurfaced, first under the guise of the infinite momentum frame (in which covariant calculations
%could be achieved more easily if the processes were viewed at the speed of light) and later as quantization on the light-front plane
%(or null plane). 
The reemergence of front-form dynamics was largely motivated by simplicity as well as physicality. The 
light-front approach has the largest stability group \cite{Leutwyler:1977vy} of any Hamiltonian theory.
%, that is light-front dynamics
%maintains the smallest number of interaction dependent generators of the Poincar\'e algebra. In highly relativistic calculations the utility is clear:
%Lorentz boosts on the light front are merely kinematical; whereas in the instant form, boosts depend on the dynamical evolution of the system. 
%The physical motivation for hard scattering processes, such as deeply inelastic scattering, was the infinite momentum frame's utility 
%in describing partonic subprocesses. 
Today the physical connection to light-front dynamics is transparent: hard scattering processes probe 
a light-cone correlation of the fields. Not surprisingly, then, many perturbative QCD applications can be treated on the light front, see e.g. 
\cite{Lepage:1980fj}. Outside this realm, physics on the light cone has been extensively developed for non-perturbative QCD 
\cite{Brodsky:1997de} as well as applied to nuclear physics \cite{Carbonell:1998rj,Miller:2000kv}.

Recently we investigated current matrix elements in the light-front Bethe-Salpeter formalism 
\cite{future}, at task originally attempted in \cite{Tiburzi:2002mn}. As an example, 
we considered the ladder approximation for the covariant kernel 
in the weak-binding limit.  We calculated $(1+1)$-dimensional electromagnetic form factors 
in this reduction scheme to demonstrate the replacement of non-wave function vertices
(as coined in \cite{Bakker:2000rd}) with contributions from higher Fock states. These higher Fock states
originate from the light-front energy pole-structure of the Bethe-Salpeter and photon vertices.  

Below, we take the model into $(3+1)$ dimensions and make clear the connection to higher Fock 
components by explicitly constructing the three-body wave function from our expressions for 
form factors. Additionally we show how the normalization of the covariant Bethe-Salpeter equation 
turns into a familiar many-body normalization in the light-front reduction. The immediate application 
of our development for form factors (in a frame where the plus-component
of the momentum transfer is non-zero) is to compute generalized parton distributions. Connection 
is again made to the Fock space representation and continuity of the distributions is put under scrutiny.
The formalism is also employed to obtain time-like form factors.
This latter application in interesting since no Fock space expansion in terms of bound
states is possible. 

The paper is organized as follows. First in section \ref{thenorm} we review the normalization of the covariant Bethe-Salpeter
wave function and then proceed to derive the light-front reduced version. This necessitates a review of the light-front reduction 
notation introduced in \cite{Sales:1999ec}. The normalization condition resembles a diagonal matrix element of a pseudo current
and has contributions from higher Fock states.  We calculate the explicit normalization condition for the ladder model at 
next-to-leading order in  perturbation theory. The connection to the familiar many-body Fock state normalization is made in 
Appendix \ref{oftopt}. Next in section \ref{gpds}, we use the reduction scheme to calculate generalized parton distributions.
We do so by using the integrand of the electromagnetic form factor calculated in an arbitrary frame
(these expressions in $(3+1)$ dimensions are collected in Appendix \ref{fff} where the leading-order bound-state 
equation for the wave function also appears).
We discuss the continuity of these distributions in terms of relations between Fock components at vanishing plus momentum. 
Connection is also made to the overlap representation of generalized parton distributions.
Application of the reduction scheme to time-like form factors and their related generalized distribution amplitudes is presented 
in section \ref{gda}. Finally we conclude with a brief summary (section \ref{summy}).    

\section{Normalization} \label{thenorm}
In \cite{Sales:1999ec} the relation between the four-dimensional Bethe-Salpeter wave function and the reduced light-front wave function was presented. 
The normalization of the reduced wave function, however, was not discussed in much detail and will be addressed below. 

Let $G(R)$ denote the two-particle disconnected propagator, where $R$ labels the total momentum. 
Thus between effective single-particle states of momenta $p$ and $k$, we have $\langle p | G(R) | k \rangle = (2\pi)^4 \delta^4(p-k) G(k,R)$, where
$G(k,R) = d(k) d(R-k)$ and the scalar single-particle propagator is
\begin{equation}
d(k) = \frac{i}{(k^2 - m^2)[1  + (k^2 - m^2) f(k^2)] + \ie}.
\end{equation} 
Here the function $f(k^2)$ characterizes the renormalized, one-particle irreducible self-interactions and for simplicity shall be ignored below.
The Bethe-Salpeter equation for the bound-state amplitude $|\Psi_R\rangle$ with mass $R^2 = M^2$ reads
\begin{equation} \label{BS}
|\Psi_R\rangle = G(R) V(R) |\Psi_R\rangle,
\end{equation}
%The reducible four-point function is defined in terms of the two-particle transition matrix $T(R)$
%\begin{equation}
%G^{(4)}(R) = G(R) + G(R) T(R) G(R),
%\end{equation}
%and has the behavior
%\begin{equation}
%G^{(4)}(R) =  - i \frac{|\Psi_{R} \rangle \langle \Psi_{R} |}{R^2 - M^2 + \ie} + \; \text{finite},
%\end{equation} 
%near the bound state pole. The $T$-matrix satisfies the usual Lippmann-Schwinger equation
%\begin{equation} \label{LS}
%T(R) = V(R) + V(R) G(R) T(R),
%\end{equation}
where $V(R)$ is the irreducible two-to-two scattering kernel (which we shall often call the potential). 
%In what follows, we remove the total momentum labels on operators since they are all identical, $R$.

%Given Eq.~\eqref{LS}, the four-point function satisfies the equation
%\begin{equation}
%G^{(4)}(R) = G^{(4)}(R) \Big( G^{-1}(R) - V(R) \Big) G^{(4)}(R),
%\end{equation}
%and hence the Bethe-Salpeter amplitude must satisfy
%\begin{equation}
%1 = \lim_{R^2 \to M^2} - i  \frac{\langle \Psi_{R}| \Big( G^{-1}(R) - V(R) \Big) |\Psi_{R}\rangle}{R^2 - M^2},
%\end{equation}
%which is necessarily finite since $|\Psi_{R}\rangle$ satisfies the bound-state
%equation: $|\Psi_{R}\rangle = G V |\Psi_{R}\rangle$. Application of l'H\^opital
%yields the covariant normalization condition
From the behavior of the reducible four-point function near the bound-state pole $R^2 = M^2$ one can deduce
the covariant normalization condition \cite{Itzykson:rh} by application of l'H\^opital's rule
\begin{equation} \label{normcov}
2 i R^\mu =  \langle \Psi_{R} | \frac{\partial}{\partial R_{\mu}} \Big( G^{-1}(R) - V(R)  \Big) |\Psi_{R} \rangle \; \Bigg|_{R^2 = M^2}.
\end{equation}
%Evaluation at $R^2 = M^2$ is understood. 
The normalization \eqref{normcov} takes the form of a diagonal 
matrix element of a pseudo current. 

The light-front reduction is performed by integrating out the minus-momentum\footnote{For any vector $a^\mu$, we define the
light-cone variables $a^\pm = \frac{1}{\sqrt{2}} (a^0 \pm a^3)$.}  dependence with the help of an auxiliary 
Green's function $\Gt(R)$. For simplicity, we denote the integration $\int \frac{d\kminus}{2\pi} \langle \kminus | \mathcal{O}(R) = \Big| \mathcal{O}(R)$.
With this notation, we will always work in $(3+1)$-dimensional momentum space for which the only sensible matrix elements of 
$\Big| \mathcal{O}(R)$ are of the form $\langle \kplus,\kperp | \;  \Big| \mathcal{O}(R) | \pminus, \pplus,\pperp \rangle$. 
The operator $\mathcal{O}(R) \Big|$ is defined similarly. To obtain light-front time-ordered perturbation theory
one chooses
\begin{equation}
\Gt(R)  =  G(R) \Big| g^{-1}(R) \Big| G(R),
\end{equation}
where $g(R) = \Big| G(R) \Big|$. This form of $\Gt$ allows for a systematic approximation scheme for the light-front energy 
poles of the Bethe-Salpeter vertex \cite{future}. Lastly one defines an auxiliary kernel $W(R)$ by \cite{Woloshyn:wm}
\begin{equation} \label{W}
W(R) = V(R) + \Big(G(R) - \Gt(R)\Big) W(R).
\end{equation}

In what follows we shall omit total four-momentum labels since they are all identically $R$. 
The normalization condition for the reduced wave function is then deduced by using the conversion \cite{Sales:1999ec}
\begin{equation} \label{324} 
| \Psi_{R} \rangle = \Bigg( 1 + \Big(G - \Gt \Big) W \Bigg) G \Big| \; |\gamma_{R} \rangle,
\end{equation}
and the definition of the reduced wave function $|\psi_R\rangle$, namely
\begin{equation} \label{psi}
|\psi_R\rangle \equiv \Big| \; | \Psi_R \rangle = g(R) |\gamma_R \rangle.
\end{equation}
Hence taking the plus component of Eq.~\eqref{normcov}
\begin{multline} \label{normlf}
2 i R^+ =  \langle \gamma_{R} | \; \Big| G \Bigg( 1 + W ( G - \Gt ) \Bigg) \\
\times \Bigg( \frac{\partial}{\partial R^-} \Big[ G^{-1} - V  \Big] \Bigg) \Bigg( 1 + (G - \Gt ) W \Bigg) G \Big| \; |\gamma_{R}\rangle.
\end{multline}
The complicated normalization condition is indicative of the effects of higher Fock space components.
To see this explicitly, we work in the ladder model in perturbation theory for which 
\begin{equation} \label{V}
V(k,p) = \frac{-g^2}{(k-p)^2 - \mu^2 + \ie}.
\end{equation}
Notice $\partial V/\partial R^\mu = 0$. 
 
Let us start with the contribution at leading order in $G - \Gt$ to the reduced wave function's normalization. 
\begin{equation} \label{normLO}
\frac{-i}{2 R^+}\langle \gamma_{R}  | \; \Big| G 
\Big( \frac{\partial}{\partial R^-} G^{-1} \Big)  G \Big| \; | \gamma_{R} \rangle = 1.
\end{equation}
To perform the integration, we note
\begin{equation}
\frac{\partial}{\partial R^-} G^{-1}(k,R) = - 2 i R^+ d^{-1}(k) (1 - x), 
\end{equation}
where we have customarily chosen $x = \kplus/R^+$. Evaluation of the integral in equation \eqref{normLO} is standard and yields
\begin{equation} \label{2to2}
N^{\text{LO}} \equiv \int \frac{dx d\kperp}{2 (2\pi)^3 x (1-x)} \psi^*(x,\kperp) \psi(x,\kperp) = 1,
\end{equation}
a simple overlap of the two-body wave function. 

To analyze the normalization to first order in $G - \Gt$, we expand equation \eqref{normlf} to first order  
\begin{equation} 
N^{\text{LO}} + \delta N  + \ldots = 1,
\end{equation}
where $N^{\text{LO}}$ is the integral appearing in \eqref{2to2} and the first-order correction arising from \eqref{normlf} is
\begin{multline} \label{deltaN}
\delta N = \frac{-i}{2 R^+} \langle \gamma_{R} | \; \Big| G \Big( \frac{\partial}{\partial R^-} G^{-1} \Big) \Big( G - \Gt \Big) V G \\ 
+  G V \Big( G - \Gt \Big) \Big( \frac{\partial}{\partial R^-} G^{-1} \Big) G \Big| \; | \gamma_{R} \rangle.
\end{multline}
The presence of $\Gt$ merely subtracts the leading-order result $N^{\text{LO}}$. 
Considering for the moment just the first term in the above equation (and omitting the subtraction $\Gt$), we have
\begin{multline}
- i \int \frac{d^4 k}{(2\pi)^4} \; \frac{d^4p}{(2\pi)^4} (1-x) \gamma^* (x,\kperp|M^2) d(k) \\
\times d(R-k)^2 V(k,p) d(p) d(R-p) \gamma (y,\pperp|M^2).
\end{multline}
The minus-momentum integrals above are similar to those considered in deriving the bound-state equation to leading order in Appendix \ref{fff}. 
The only difference is the double pole due to the extra propagator $d(R-k)$. With $x = \kplus/R^+$ and $y = \pplus/R^+$, for $x>y$
we avoid picking up the residue at the double pole and the result is the same as in Eq.~\eqref{normLO} after using the bound-state equation.
This term is then subtracted by the $\Gt$ term in equation \eqref{deltaN}. On the other hand, when $x<y$ we pick up the residue at the double
pole. Part of the residue is subtracted by the $\Gt$ term; the other half depends on $\partial V(k,p)/\partial k^-$. The second term in Eq.~\eqref{deltaN}
is evaluated identically up to $\{k \leftrightarrow p\}$.  Now combining the two terms and their relevant 
$\theta$ functions, we can rewrite the result using the explicit form of the one-boson exchange potential Eq.~\ref{OBE}, namely
\begin{widetext}
\begin{equation} \label{nonN}
\delta N = \int \frac{dx d\kperp}{2(2\pi)^3 x (1-x)} \; \frac{dy d\pperp}{2 (2\pi)^3 y (1-y)} \psi^*(x,\kperp) \Bigg( - \frac{\partial}{\partial M^2}      
V(x, \kperp; y, \pperp |M^2) \Bigg)  \psi(y, \pperp).  
\end{equation} 
\end{widetext}
Thus although the covariant derivative's action on the potential vanishes, we can manipulate the correction to the normalization into the form of a 
derivative's action on the light-front, time-ordered potential. With this form, we can compare to the familiar nonvalence probability discussed in Appendix \ref{oftopt} (in the frame where $R^+ = R^- = M/\sqrt{2}$ with $M/\sqrt{2}$ as the eigenvalue of the light-front Hamiltonian, denoted $\pminus$ in Eq.~\eqref{quspace}). 

Here we have seen that the normalization of the light-cone wave function includes effects from higher Fock states. Since this normalization
condition stemmed from a diagonal matrix element of a pseudo current, we should not be surprised that matrix elements of the electromagnetic 
current, when treated in this reduction scheme, pick up contributions from higher Fock states. 
These higher Fock contributions appear explicitly and are the subject of the next section.

\section{Application to GPDs} \label{gpds}
Having worked through matrix elements of the electromagnetic current in a frame where $\Delta^+ \neq 0$ \cite{future}, 
we can now make the connection to generalized parton distributions (GPDs). These distributions, which in some sense 
are the natural interpolating functions between form factors and quark distribution functions, turn up in a variety of hard exclusive processes, 
e.g.  deeply virtual Compton scattering, wide-angle Compton scattering and the electro-production of mesons \cite{Muller:1994fv}. The scattering amplitude for these processes factorizes into a convolution
of a hard part (calculable from perturbative QCD) and a soft part which the GPDs encode. Since light-cone correlations are probed in these 
 hard processes, the soft physics has a simple interpretation and expression in terms of light-front wave functions \cite{Brodsky:2001xy}.
In this section, we cast our results for form factors \cite{future} in the language of GPDs
and the light cone Fock space expansion. The $(3+1)$-dimensional expressions for form factors are presented in Appendix \ref{fff}. 
Additionally one can obtain these results directly from time-ordered perturbation theory using two-body projection operators 
as explicated in Appendix \ref{oftopt}. 

The GPD for our meson model is defined by a non-diagonal matrix element of bilocal field operators
\begin{equation} \label{bilocal}
F(x, \zeta, t) = \int \frac{dy^-   }{4\pi} e^{i x P^+ y^-}
\langle \Psi_{P^\prime} | \; q(y^-)   i \overset{\leftrightarrow}\partial{}^+ q(0) \; | \Psi_{P} \rangle,
\end{equation}
where $q(x)$ denotes the quark field operator and $\overset{\leftrightarrow}\partial{}^\mu = \overset{\rightarrow}\partial{}^\mu - 
\overset{\leftarrow}\partial{}^\mu$. 
%The unusual pre-factor of $x/(2x - \zeta)$ corresponds to having scalar quarks 
%and derivative coupling at the quark-photon vertex. 
Comparing to the current matrix element $J^\mu$ in Appendix \ref{fff}, the definition of the GPD leads immediately to the sum rule
\begin{equation} \label{sumrule}
\int \frac{dx}{1 - \zeta/2} F(x, \zeta, t) = F(t).
\end{equation}
Hence one can calculate these distributions from the integrand of the form factor. In this way, the light-cone correlation defined in 
Eq.~\eqref{bilocal} has a natural description in terms of light-front time-ordered perturbation theory, e.g. for $x>\zeta$ the 
relevant graphs contributing to the GPD are in Figure \ref{ftri} and \ref{ftri2}, and those for $x<\zeta$ are in Figure \ref{fZZZ}. 
%%%%%Also note, the $\zeta$ dependence disappears after the $x$ integration present in the sum rule due to Lorentz invariance. 
%We were aware of this in section \ref{current} when we opted not to choose a frame in which $\Delta^+ = 0$ for the form factor when we knew at the end of the day $F(t)$ would be a function of $t = \Delta^2$ alone.

\subsection{Continuity}
Conversion of the contributions to the form factor into GPDs is straightforward using Eq.~\eqref{sumrule}; we merely 
remove $- 2 i P^+$ and $\int dx$ from Eqs.~(\ref{FFLO}-\ref{fs2}). In order for the deeply virtual Compton scattering 
amplitude to factorize into hard and soft pieces (at leading twist), 
the GPDs $F(x,\zeta,t)$ must be continuous at $x = \zeta$. Maintaining continuity at the crossover is more pressing because  
experiments which measure the beam-spin asymmetry are limited to the crossover \cite{Diehl:1997bu}. The leading-order expressions are continuous. This 
is easy to see since the contribution for $x < \zeta$ is identically zero. The valence contribution for $x>\zeta$ is a convolution of wave functions
one of which is $\psi^*(x^\prime,\ldots)$ which is probed at the end point since $x^\prime \equiv \frac{x - \zeta}{1-\zeta} \to 0$. 
From the bound-state equation Eq.~\eqref{wavefunction}, 
we see the two-body wave function vanishes quadratically at the end points. Taking into account the overall weight $x^{\prime -1}$, the 
valence piece vanishes linearly at the crossover. At leading order then, continuity is maintained at the crossover, while the derivative
is discontinuous. Working only in the valence sector, valence quark models will never be of any use to beam-spin asymmetry measurements 
since the value at the crossover requires one wave function to be at an end point. In the three-body bound state problem (e.g. the nucleon), 
%the wave function again necessarily vanishes (in our conventions); however, 
the valence GPD will vanish only if the three-body interaction 
is non-singular at the end points (which is physically reasonable and perturbatively true). 

Let us now check the next-to-leading order contributions to the GPD for continuity. First we shall deal with the term stemming from iterating 
the Bethe-Salpeter equation for the initial state \eqref{is1} (see diagram $B$ of Figure \ref{ftri2}).
 Since there is no Z-graph generated from iterating the initial state, we expect this contribution to vanish. 
Looking at the expression, we see again 
$\psi^*(x^\prime,\ldots)/x^\prime$ which vanishes linearly as $x \to \zeta$. Moreover there are the interaction terms: 
$D(y,\pperp;x,\kperp|M^2) $ which is finite as $x \to \zeta$, and 
\begin{equation}
D(\yp,\ppp;\xp,\kpp|M^2) \overset{x\to \zeta}{=} \frac{-\xp}{(\kperp + \mathbf{\Delta}^\perp)^2 + m^2}, 
\end{equation}
which vanishes at the crossover. Thus not only does the initial-state iteration term vanish at the crossover, its derivative does so as well.

Now we investigate the Born terms \ref{bt1} (see diagram $A$ of Figure \ref{ftri2})  and \ref{bt2} (see diagram $D$ of Figure \ref{fZZZ}) at the crossover. Approaching $\zeta$ from above \eqref{bt1}, we 
%encounter
%two interactions $\tilde{V}(\kminus_{a}, \ldots)$ which is finite as above and $\tilde{V}(\kminus_c, \ldots)$ which vanishes linearly (enough
%to cancel the weight $(x - \zeta)^{-1}$. Thus 
have the finite contribution at the crossover
\begin{widetext}
\begin{equation}\label{bcross}
F(\zeta, \zeta, t)^{\text{Born}} =  \int \frac{d\kperp dy d\pperp}{(16\pi^3)^2  y (1-y)\yp }  
\psi^*(y^\prime, \ppp) \frac{g^2 \theta(y - \zeta)/(y - \zeta) }{(\kperp + \mathbf{\Delta}^\perp)^2 + m^2} 
D(y,\pperp;\zeta,\kperp|M^2) \psi(y, \pperp). 
\end{equation} 

On the other hand, approaching the crossover from below \eqref{bt2} we have to deal with singularities as $x^{\prime\prime} = x/ \zeta \to 1$. Writing out
the propagator for the quark-antiquark pair heading off to annihilation, we see
\begin{equation} \label{qqbarprop}
\dw(x^{\prime\prime}, \kppp | t ) \to -  \frac{1 - x^{\prime\prime}}{(\kperp + \mathbf{\Delta}^\perp)^2 + m^2}.
\end{equation}
This linear vanishing cancels the weight $(1 - x^{\prime\prime})^{-1}$. Taking the limit $x \to \zeta$ then produces
equation \eqref{bcross} and thus the Born terms are continuous. 

Lastly we must see how the final-state iteration terms match up at the crossover. 
%Having spelled out the Born terms, the final-state
%terms follow simply once we note $\tilde{V}(\kminus_{a}, x^\prime, \kperp + \mathbf{\Delta}^\perp; y^\prime, \qperp)$ is finite as $x^\prime \to 0$ and
%\begin{equation}
%\theta(y^\prime - x^\prime) \tilde{V}(\kminus_{c}, x^\prime, \kperp + \mathbf{\Delta}^\perp; y^\prime, \qperp) \to \frac{-g^2 x^\prime}{y^\prime} 
%\frac{1}{(\kperp + \mathbf{\Delta}^\perp)^2 + m^2},
%\end{equation}  
%which vanishes as $x$ goes to $\zeta$. 
Using Eq.~\eqref{fs1} to approach $\zeta$ from above (see diagram $C$ of Figure \ref{ftri2}), we have the contribution
\begin{equation}\label{fcross}
F(\zeta, \zeta, t)^{\text{final}} = \int \frac{d\kperp d\yp d\ppp}{(16\pi^3)^2 (1-\zeta)\yp (1-\yp) } 
\psi^*(y^\prime, \ppp) \frac{g^2/y }{(\kperp + \mathbf{\Delta}^\perp)^2 + m^2} 
D(y,\pperp;\zeta,\kperp|M^2) \psi(\zeta, \kperp).
\end{equation}
Approaching $\zeta$ from below (see diagram $E$ of Figure \ref{fZZZ}), we utilize equation \eqref{qqbarprop} in taking the limit of \eqref{fs2}. The result is \eqref{fcross}
and hence we have demonstrated continuity to first order, i.e.
\begin{equation} \label{contequ}
F(\zeta,\zeta,t) = F(\zeta,\zeta,t)^{\text{Born}} +   F(\zeta,\zeta,t)^{\text{final}}
\end{equation}
no matter how we approach $x = \zeta$. 

\subsection{Fock space representation}
We now write the GPDs in terms of Fock component overlaps. In the diagonal overlap region $x > \zeta$ this will be a mere rewriting 
of our results, while there is a subtlety for the non-diagonal overlaps. To handle the zeroth-order term \eqref{FFLO}, we 
define the two-body Fock component as
\begin{equation} \label{twofock}
\psi_{2}(x_{1},\mathbf{k}^\perp_1, x_{2}, \mathbf{k}^\perp_2) = \frac{1}{\sqrt{ x_{1} x_{2} }} \psi(x_{1}, \mathbf{k}^\perp_{\text{rel}}), 
\end{equation}
noting that the relative transverse momentum can be defined as $\mathbf{k}^\perp_{\text{rel}} = x_{2} \mathbf{k}^\perp_{1} - x_{1} \mathbf{k}^\perp_{2}$. 
In terms of Eq. \eqref{twofock}, the GPD appears as 
\begin{equation} \label{222}
F(x, \zeta, t)^{\text{LO}} = \frac{\theta(x - \zeta)}{\sqrt{1-\zeta}} \int [dx]_{2} [d\kperp]_{2} \sum_{j = 1,2} \delta(x - x_{j})
  \psi_{2}^*(x^{\prime}_{i}, \mathbf{k}^\prime_{i}{}^\perp) \frac{2 x_j - \zeta}{\sqrt{ x^{\prime}_{j} x_j }}  \psi_{2}(x_{i},\mathbf{k}_{i}^\perp),  
\end{equation}
where the primed variables are given by
\begin{equation} \label{primed}
\begin{cases}
x^\prime_{i} = \frac{x_{i}}{1-\zeta}\\
\mathbf{k}^\prime_{i}{}^\perp  = \mathbf{k}_{i}^\perp - x^\prime_{i} \mathbf{\Delta}^\perp, \; \text{for} \; i \neq j
\end{cases}
\begin{cases}
x^\prime_{j} = \frac{x_{j} - \zeta}{1 - \zeta}\\
\mathbf{k}^\prime_{j}{}^\perp = \mathbf{k}_{j}^\perp + (1 - x^\prime_{j}) \mathbf{\Delta}^\perp
\end{cases}
\end{equation}
and the integration measure is given by
\begin{align}
[dx]_{N} & = \prod_{i = 1}^{N} dx_{i} \; \delta \Big( 1 - \sum_{i = 1}^{N} x_{i} \Big)\\
[d\kperp]_{N} & = \frac{1}{[2(2\pi)^3]^{N-1}} \prod_{i=1}^N d\mathbf{k}^\perp_{i} \; \delta \Big( \sum_{i=1}^N \mathbf{k}^\perp_{i} \Big).
\end{align}
Notice the sum over transverse momenta in the delta function is zero since our initial meson has $\mathbf{P}^\perp = 0$.
The sum over $j$ in Eq. \eqref{222} produces the overall factor of two for our case of equally massive (equally charged) constituents. 

To cast the next-to-leading order expressions for $x > \zeta$ in terms of diagonal Fock space overlaps, we must write out the three-body Fock component.
Looking at the diagrams in Figure \ref{ftri2}, it is constructed from the two-body wave function
\begin{multline} \label{threefock}
\psi_{3} (x_{i},\mathbf{k}_{i}^\perp) = g \frac{2 (2\pi)^3}{\sqrt{x_{1} x_{2} x_{3}}} 
\int [dy]_{2} [d\pperp]_{2} \Bigg[ \theta(y_{1} - x_{1}) x_{3} \delta(y_{2} - x_{3}) 
\delta(\mathbf{p}^\perp_{2} - \mathbf{k}^\perp_{3}) D(y_1,\mathbf{p}^\perp_1;x_1,\mathbf{k}^\perp_1|M^2)
\\ + \theta(y_2 - x_3) x_1 \delta(y_{1} - x_{1}) \delta(\mathbf{p}^\perp_{1} - \mathbf{k}^\perp_{1}) 
D(y_2,\mathbf{p}^\perp_2;x_3,\mathbf{k}^\perp_3|M^2)
\Bigg] \frac{\psi_{2}(y_{j},\mathbf{p}_{j}^\perp)}{\sqrt{y_{1} y_{2}}},
\end{multline}
where $i$ runs from one to three and the label $j$, which stems from the integration measure, runs from one to two. 
%The awkward looking factor of 
%$\sqrt{x_2/g^2}$ arises from our definition of $\tilde{V}$. It serves to remove a factor of $g$ since only one interaction is needed to place us in the 
%three-body subspace and when combined with the prefactor in $\tilde{V}$ produces the factor $1/\sqrt{x_{2}}$ which then symmetrically appears
%with the overall constant in $\psi_{3}$.  
We discuss how to obtain this three-body wave function directly from time-ordered perturbation theory in 
Appendix \ref{oftopt}. Using $\psi_3$ in Eq. \eqref{threefock}, the terms in the GPD at first 
order in the weak coupling can then be written compactly as
\begin{equation} \label{323}
F(x,\zeta,t)^{\text{NLO}} = \frac{\theta(x - \zeta)}{1 - \zeta} \int [dx]_3 [d\kperp]_{3} \sum_{j = 1,3} 
\delta( x - x_{j}) \psi_{3}^*(x_{i}^\prime,\mathbf{k}_{i}^\prime{}^\perp) \frac{2 x_j - \zeta}{\sqrt{ x^{\prime}_{j} x_j }}
\psi_{3}(x_{i},\mathbf{k}^\perp_i).
\end{equation}
\end{widetext}
One can verify that the diagrams in Figure \ref{ftri2} are generated by \eqref{323}. 
%Note well that our definition of $z$
%as a momentum fraction with respect to the final plus momentum $P^\prime{}^+$ already includes the factor of $(1-\zeta)^{-1}$ with respect 
%to $P^+$. 
Additionally there is a fourth diagram generated by Eq. \eqref{323} which does not appear in the figure. This missing diagram is characterized
by the spectator quark's one-loop self interaction and is absent since we have ignored $f(k^2)$ and the scale dependence of light-cone wave 
functions.\footnote{Recently the scale dependence of light-cone Fock components has been investigated \cite{Burkardt:2002uc}.}
%Thus in performing calculations in perturbation theory, we should also use
%the one-loop result for the renormalized self interactions $f(k^2)$ appearing in the spectator's propagator. As far as our development is concerned, 
%we imagined this contribution tacked onto Figure \ref{ftri}, while it appears now as part of the three-to-three Fock component overlaps. 
The absence of this diagram does not affect continuity at the crossover. The missing diagram vanishes at $x = \zeta$ since the 
final-state wave function is $\psi^*(x^\prime,\ldots)$. Lastly we note the above Fock component overlaps satisfy the positivity
constraint for a composite scalar composed of scalar constituents \cite{Tiburzi:2002kr}. 

Now we must come to terms with the non-diagonal overlap region, $x < \zeta$. At first order, the diagrams of Figure \ref{fZZZ}
correspond to four-to-two Fock component overlaps. We have been cavalier about time ordering, however. The expressions  
Eqs. \ref{bt2} and \ref{fs2} do not correspond to time-ordered graphs. Both terms contain a product of time-ordered propagators:
one for the two quarks leading to the final-state vertex and another for the quark-antiquark pair heading off to annihilation. But for an interpretation
in terms of a four-body wave function, all four particles must propagate at the same time. This is a subtle issue as a graph containing the product 
of two independently time-ordered pieces (where one leads to a bound-state vertex) corresponds to a sum of infinitely many time-ordered graphs. 
It is easiest to write out the terms of concern in terms of the propagators' poles. The quark, anti-quark heading to annihilation
have propagators $d(k)$ and $d(k+\Delta)$ and poles we label $k^-_a$ and $k^-_c$, respectively. The remaining propagators of interest
$d(p^\prime)$ and $d(P^\prime - p^\prime)$ have poles $p^-_a$ and $p^-_b$, respectively.
We can then manipulate as follows
\begin{multline} 
%- (2 P^+)^2 \zeta (1-\zeta) \dw(\sigma, \kperp + \sigma \mathbf{\Delta}^\perp |t) \dw(z, \qperp - z \mathbf{\Delta}^\perp|M^2) & = 
\frac{1}{k^-_{c} - k^-_{a}} \; \frac{1}{p^-_{b} - p^-_{a}} \\
	=  \frac{1}{p^-_{b} - p^-_{a} + k^-_{c} - k^-_{a}} \; \Bigg( 
\frac{1}{p^-_{b} - p^-_{a}} + \frac{1}{k^-_{c} - k^-_{a}} \Bigg).   \label{trickery}
\end{multline}
In this form, we have produced the correct energy denominator for the instant of light-front time where four particles are propagating. 
Multiplying this denominator by the three-body wave function yields the four-body wave function (up to constants).
This is the part of the four-body wave function relevant for GPDs (there are additional pieces for two-quark, two-boson states, see Appendix 
\ref{oftopt}).  
In the resulting sum \eqref{trickery}, the first term will produce the two-body wave function for the final state and we will have a
genuine four-to-two overlap. We do not write this out explicitly.

The second term in Eq. \eqref{trickery}, however, contains again the propagator for the pair heading to annihilation. Using the light-front 
Bethe-Salpeter equation for the vertex (which contains infinitely many times) we can introduce a factor of the time-ordered interaction. 
The resulting product of independent time orderings can again be manipulated as in Eq. \eqref{trickery}. The result produces another overall four-body
denominator which contributes to the four-body Fock component of the initial state. Since we iterated the interaction, however, this new contribution
is no longer at leading order and can be neglected. Thus the second term in \eqref{trickery} does not contribute at this order.

Having manipulated the GPDs into non-diagonal overlaps for $x < \zeta$, we must wonder if continuity at the crossover is still maintained. In the limit 
$x \to \zeta$ the light-front energy of the struck quark goes to infinity. %(since it has vanishing plus momentum). 
Consequently $k^-_{c}$, 
which contains this on shell energy, is infinite and dominates the four-body energy denominator. This is identical to the reasoning
in Eqs.~\ref{bt2} and \ref{fs2} where instead of the four-body denominator, we have $\dw(x^{\prime\prime}, \kperp + x^{\prime\prime} 
\mathbf{\Delta}^\perp |t)$ which is dominated by $k^-_{c}$ at the crossover. Either way, 
we arrive at the expressions found above for the crossover \eqref{bcross} and \eqref{fcross}.

Having cast our expressions for generalized parton distributions in terms of the Fock components, 
%generated to first order in the weak coupling
we can enlarge our understanding of the sum rule and continuity at the crossover. Both must deal with the relation between higher Fock components. The way 
$\zeta$-dependence disappears from \eqref{sumrule} mandates a relation between the diagonal and non-diagonal Fock component overlaps that make up the 
GPD. The relation between Fock components must follow from the field-theoretic equations of motion. Continuity itself is a special
case of the relation between Fock components, specifically at the end points. Above we have seen our expressions are continuous (and non-vanishing)
at the crossover and explicitly that the three- and four-body components match at the end point (where $x - \zeta = 0$). This weak binding model for behavior 
at the crossover is a simple example of the relations between Fock components at the end points (see also \cite{Antonuccio:1997tw}). More general 
relations must be permitted from the equations of motion to guarantee Lorentz covariance (e.g. in the structure of the Mellin moments of the GPDs, of which 
the sum rule is a special case). Here, of course, Lorentz symmetry is broken. Infinitely many light-cone time-ordered graphs are needed
in the reduced kernel to reproduce the covariant one-boson exchange \eqref{V}. Thus exactly satisfying polynomiality requires not only infinitely many 
exchanges in the kernel but contributions from infinitely many Fock components. It should be possible, however, to show how
the sum rule and polynomiality are improved order by order. 

 \section{Application to GDAs} \label{gda}
Below we study time-like form factors to demonstrate the versatility of this approach and make the connection to the generalized
 distribution amplitudes (GDAs) for this model. Analogous to GPDs, GDAs encode the soft physics of two-meson production and can thus be thought of 
 as  crossed versions of the GPDs. The GDAs enter in convolutions for various two-meson production amplitudes \cite{Diehl:1998dk}. 
These distribution functions as well as time-like form factors are a theoretical challenge for light-front dynamics, 
since there is no direct decomposition in terms of meson Fock components alone. 
Furthermore, we shall see the leading-order expressions are non-valence contributions 
(which necessarily excludes a description in terms of most constituent quark models). 

\begin{figure}
\begin{center}
\epsfig{file=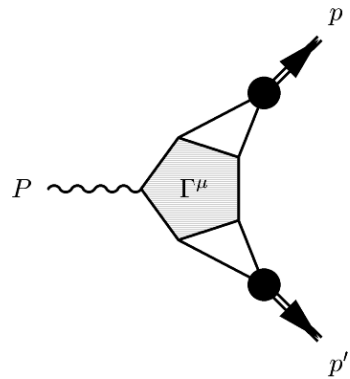}
\caption{The triangle diagram for the time-like, pion form factor}
\label{ftimetri}
\end{center}
\end{figure}

The time-like form factor $F(s)$ for our model meson is defined by (see Figure \ref{ftimetri})
\begin{equation}
\langle \Psi_{p} \; \Psi_{p^\prime} | \; \Gamma^\mu \; | 0 \rangle = - i (p - p^\prime)^\mu F(s),
\end{equation}
where $s = (p + \pp)^2$ is the center of mass energy squared. Now define $P^\mu = p^\mu + \pp{}^\mu$ and $\zeta = \pplus / \Pplus$. We can work out the kinematics of this reaction in a frame where $\mathbf{P}^\perp = 0$ 
\begin{align}
P^- & = \frac{s}{2 P^+} \notag \\
p^- & = \frac{(1 - \zeta) s}{2 \Pplus} \notag \\
\mathbf{p}^\perp{}^2 & = s (1 - \zeta) \zeta - M^2,
\end{align}
where $M$ is the meson mass. 

Similar to GPDs, the GDA for our model has a definition in terms of a non-diagonal matrix element of bilocal field operators
\begin{equation}
\GDA = \int \frac{dx^-}{2 \pi}  e^{i z \Pplus x^-} \langle \Psi_{p} \; \Psi_{\pp} | \; q(x^-) i \
\overset{\leftrightarrow}\partial{}^+ q(0) \; | 0 \rangle.
\end{equation}
Such a definition of the GDA leads directly to a sum rule for the time-like form factor
\begin{equation} \label{sumtime}
\int \frac{dz}{2 \zeta - 1} \GDA = F(s),
\end{equation}
and hence a means to calculate $\Phi$ from the integrand of the time-like form factor. 

Taking the appropriate residues of the five-point function, we arrive at Fig. \ref{ftimetri} for the time-like form factor. Keeping only the leading order piece of the electromagnetic vertex $\Gamma^\mu$, we have
\begin{multline}
\GDA = i \Pplus (2z -1) \int \frac{d\kminus d\kperp}{(2\pi)^4} 
\\ \times \gamma^*( \zpp, \kperp - \zpp \pperp | M^2 ) G(k,p) d^{-1}(k-p)
\\ \times  G(P-k, \pp)  \gamma^*\big( \zp, \kperp  - (1- \zp) \pperp \big| M^2 \big), 
\end{multline}
where we have made use of $z = \kplus/ \Pplus$, $\zp = \frac{z - \zeta}{1 - \zeta}$ and $\zpp = z / \zeta$. 
%Since this term carries no overall factor of the coupling, we must have only wave function vertices (knowing well that any apparent non-wave function vertices will be subtracted in the complete next-order expression). 
Recall $\gamma(x|M^2) \propto \theta[ x(1-x)]$. 
This translates to: $0< \zpp < 1$ and $0< \zp < 1$, and hence we do not pick up a contribution at zeroth order in the coupling. 

To work at first order, we pick up three terms analogous to those in Appendix \ref{fff}. We denote these as $\delta J^\mu_{\gamma}, \delta J^\mu_{p}$ and
$\delta J^\mu_{\pp}$. The Born term for the three-point electromagnetic vertex $\delta J^\mu_{\gamma}$ is quite simple. 
%Any diagrams contributing to this term must contain an overall power of the weak coupling since the boson can never be reduced into either final-state vertex. Thus we must restrict the plus-momentum fractions of the two final-state vertices to be between zero and one. 
For the same reason as the zeroth-order result, the restriction of $\gamma(x|M^2) \propto \theta[x(1-x)]$ and momentum conservation
force the contribution $\delta J^\mu_{\gamma}$ to vanish. This leaves us to consider only diagrams that arise from iteration of the 
Bethe-Salpeter equation of either final-state meson.

Considering first the term $\delta J^\mu_{p}$, we have the contribution to the GDA
%\begin{widetext}
\begin{multline} \label{blah}
\Phi_{p}(z,\zeta,s)  =  i \Pplus (2 z -1) \int \frac{d\kminus d\kperp}{(2\pi)^4} \; \frac{d^4q}{(2\pi)^4}
\\ \times \gamma^*(\ypp, \qperp - \ypp \pperp) G(q,p) V(q,k) G(k,p) G(P - k, \pp) 
\\ \times d^{-1}(k-p) \gamma^*(\zp, \kperp - (1-\zp) \pperp), 
\end{multline}
where we have chosen to abbreviate $y = \qplus/ \Pplus$ and hence the label $\ypp = y / \zeta$. We have customarily omitted the subtracted term containing $\Gt$, which is zero because there is no leading-order term to subtract.
%we know it removes the non-wave function vertices present in Eq. \ref{blah}.
%which have the same form as those encountered at zeroth order. Thus we can entirely omit any non-wave function vertices, knowing well they will be canceled. 
Requiring wave function vertices mandates $0<\ypp<1$ and $\zeta < z < 1$.
%, which will be of incredible aid in evaluating the integrals. 

%With these restrictions, the $\qminus$ integral can be performed similar to equation \eqref{firsto} and the subsequent $\kminus$ integration resembles that in section \ref{current} but here all the imaginary parts' signs are fixed. 
Thus $\Phi_{p}$ produces one contribution to the GDA 
%from $\delta J^\mu_{p}$
\begin{widetext}
\begin{multline} \label{jp}
\Phi_{p}(z,\zeta,s)  = \frac{\theta(z - \zeta)}{(16\pi^3)^2 \zeta} \int \frac{d\kperp dy d\qperp (2 z - 1)}{z (1-z) \zp \ypp(1-\ypp)} 
 \dw(z, \kperp |s)  \frac{g^2 \theta(z - y)}{z - y} 
\\ \times  D(z,\kperp;y,\qperp|s) \psi^*(\zp, \kperp - (1-\zp) \pperp) \psi^*(\ypp, \qperp - \ypp \pperp).  
\end{multline}
%where for clarity we spell out the form of the interaction
%\begin{equation} 
%\frac{g^2 \theta(z - \zeta)}{2 \pplus \Big( \frac{z}{\zeta} - w  \Big)} \tilde{V} \Big(\kminus_{b} , \frac{z}{\zeta}, \kperp; w, \qperp\Big)^{-1}  = 
%P^- - q^-_{\text{on}} - \frac{(\kperp - \qperp)^2 + \mu^2}{2 (\kplus - \qplus)} - (P - k)^-_{\text{on}}.
%\end{equation}
As a contribution to the time-like form factor, we can interpret Eq. \eqref{jp} as the time-ordered diagram $b$ of Figure \ref{fgda}. 

\begin{figure}
\begin{center}
\epsfig{file=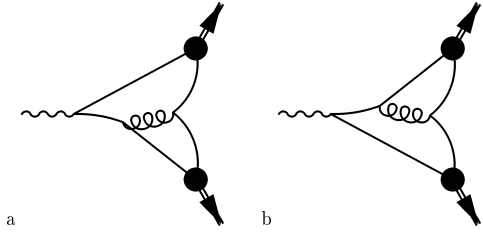,width=2.5in}
\caption{Leading order diagrams for the time-like, pion form factor.}
\label{fgda}
\end{center}
\end{figure}

At first order in the weak coupling, we have one term remaining to consider $\delta J^\mu_{\pp}$. Again omitting the 
superfluous subtraction of $\Gt$, we have
\begin{multline}
\Phi_{p^\prime}(z,\zeta,s) = i \Pplus (2 z -1) \int \frac{d\kminus d\kperp}{(2\pi)^4} \; \frac{d^4 q}{(2 \pi)^4} \gamma^*(\zpp, \kperp - \zpp \pperp | M^2) G(k,p) 
d^{-1}(p-k) \\ \times G(P - k , \pp) V(P-k, P-q) G(P-q, \pp) \gamma^*(\yp, \qperp - (1- \yp ) \pperp|M^2), 
\end{multline}
with $\yp = \frac{y - \zeta}{1-\zeta}$.
%Since we are removing the crossed interactions by hand, we restrict 
Both $\yp$ and $z$ are restricted: $0 < \yp < 1$ and $0 < z < \zeta$, and  
%The signs of all poles' imaginary parts
%are now fixed and the integration proceeds just as many above. 
%We thus arrive at 
the final contribution to the GDA is
\begin{multline} \label{jpp}
\Phi_{p^\prime}(z,\zeta,s) = \frac{\theta(\zeta - z)}{(16\pi^3)^2 \zeta} \int \frac{d\kperp d\yp d\qperp (2 z - 1)}{(1-z) \zpp (1-\zpp) \yp (1-\yp)} 
\dw(z, \kperp | s)  \frac{g^2 \theta(y - z)}{y-z} 
\\ \times  D(y,\qperp;z,\kperp|s) \psi^*(\zpp, \kperp - \zpp \pperp) \psi^*(\yp, \qperp + \yp \pperp), 
\end{multline} 
%where we have defined $\sigma = z / \zeta$ and the interaction is
%\begin{equation}
% \frac{g^2 \theta(\zeta - w) }{2 \pp{}^+ \Big( \frac{1 - z}{1 - \zeta} - w \Big)} \tilde{V}(\kminus_{a}, 1 - z, -\kperp; w, \qperp)^{-1} = 
%P^- - \kminus_{\text{on}} - [(P - k) - q]^-_{\text{on}} - q^-_{\text{on}} .
%\end{equation}
\end{widetext}
In this form we recognize this contribution as diagram $a$ of Figure \ref{fgda}.

Having found the leading non-vanishing contribution to the GDA namely $\Phi = \Phi_{p} + \Phi_{p^\prime}$, we observe that the higher Fock components derived in section \ref{gpds} (as well as in Appendix \ref{oftopt}) do not fit naturally into \eqref{jp} or \eqref{jpp}. One needs a Fock space expansion for the photon wave function in order to have an expression for the GDA in terms of various Fock component overlaps. With the expressions derived for the GDA we can use Eq. \eqref{sumtime} to obtain the time-like form factor. 

\section{Summary} \label{summy}
Above we have investigated various applications of the light-front reduction of current matrix elements. 
First we considered the normalization of the light-front wave function in the reduction formalism
deriving Eq.~\eqref{normlf} from the covariant normalization Eq.~\eqref{normcov}. The complicated form of
the reduced normalization was linked to effects of higher Fock components (which we illustrated
by using the ladder model \eqref{V} in perturbation theory). Using the explicit form of the leading-order kernel, 
we were able to derive Eq.~\eqref{nonN}, which is the familiar many-body normalization condition \eqref{quspace}. 

In Appendix \ref{fff}, we reviewed the derivation of the form factor at next-to-leading order
in the $(3+1)$-dimensional ladder model. These expressions were then converted into the GPD
for the model. Continuity of these distributions at the crossover (where the plus momentum of 
the struck quark is equal to the plus component of the momentum transfer) was explicitly demonstrated,
\emph{cf} Eq.~\eqref{contequ}. Connection was made to the overlap representation of GPDs
by constructing the three-body wave function to leading order in perturbation theory. 
As a check on our results, we also reviewed the construction of higher Fock states from the valence sector in
old-fashioned time-ordered perturbation theory (Appendix \ref{oftopt}). The derived
overlaps Eqs.~\eqref{222} and \eqref{323} satisfy the relevant positivity constraint. 
The non-vanishing of the GPDs at the crossover could then be tied to higher Fock components,
specifically at vanishing plus momentum, and are hence essential for any phenomenological modeling of these distributions.  
This rewriting allowed us to understand how continuity arises perturbatively from the small-$x$ behavior of Fock state wave functions. 
In perturbation theory, the diagonal valence overlap vanishes at the crossover, 
while the higher Fock component overlaps do not. In general 
the $n$-to-$n$ overlap matches up with the $(n+1)$-to-$(n-1)$ overlap at the crossover due to the
dominance of the rebounding quark's infinite energy. The same seems to be true perturbatively for a three-body 
bound state due to the nature of the kernel.

Unfortunately issues involving Lorentz invariance (such as the sum rule for the electromagnetic form
factor and the polynomiality constraints) are left untouched. To maintain covariance 
one would need infinitely many time-ordered exchanges in the kernel as well as 
infinitely many Fock components. It should be possible, however, to 
understand perturbatively how the $\zeta$-dependence disappears from Eq.~\eqref{sumrule}. 
This requires further relations between Fock components and these should be afforded
by the field-theoretic equations of motion. 

Lastly we considered application of the reduction formalism for currents
to the time-like form factor. We did so by calculating the ladder model's 
GDA Eqs.~(\ref{jp}-\ref{jpp}), which is related to the time-like form factor via the sum rule in Eq.~\eqref{sumtime},
systematically in perturbation theory. This is in contrast to the non-existent Fock space expansion for these types of 
processes. 

With the formalism explored here, one could use phenomenological Lagrangian based models to explore both generalized
parton distributions and generalized distributions amplitudes within the light-front framework. Such an investigation
is interesting not only for testing phenomenological models, but also for anticipating problems for approximate non-perturbative
solutions for the light-cone Fock states. Nonetheless more model studies are warranted before truly realistic calculations
can be pursued. 

\begin{acknowledgments}
We thank M.~Diehl for enthusiasm, questions and critical comments.
This work was funded by the U.~S.~Department of Energy, grant: 
DE-FG$03-97$ER$41014$.  
\end{acknowledgments}

\appendix

\section{Wave functions and form factors in $(3+1)$ dimensions} \label{fff}
In this Appendix we collect results relevant above for wave functions and form factors in the $(3+1)$-dimensional ladder model Eq.~\eqref{V}.

\subsection{Wave functions}
Using the Bethe-Salpeter equation \eqref{BS} and the definition of the light-cone wave function $|\psi_R\rangle$ \eqref{psi}
we have the light-cone bound-state equation 
\begin{equation} \label{bse}
|\psi_{R}\rangle = g(R) w(R) |\psi_R \rangle,
\end{equation}
where $w(R)$ is the reduced auxiliary kernel
\begin{equation}
w(R) = g^{-1}(R) \Big| G(R) W(R) G(R) \Big| g^{-1}(R),
\end{equation}
with $W(R)$ defined in Eq.~\eqref{W}.

\begin{figure}
\begin{center}
\epsfig{file=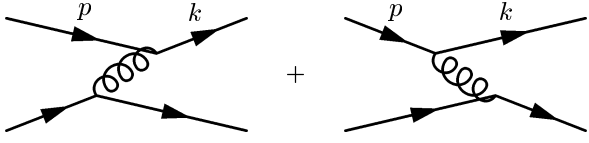}
\end{center}
\caption{Diagrammatic representation of the one-boson exchange potential $V$ appearing in Eq. \eqref{OBE}.}
\label{fOBEP}
\end{figure}

To leading order in $G - \Gt$, one calculates $w(R)$ for the ladder model to be:
%\begin{widetext}
\begin{multline} \label{OBE}
V(x,\kperp;y,\pperp|M^2) \equiv - \langle \; xR^+,\kperp | \; w(R) \; | \; yR^+,\pperp \rangle \\
			 = \frac{g^2}{x-y}  \Big[ \theta(x-y) D(x,\kperp;y,\pperp|M^2) \\
- \big\{(x,\kperp) \longleftrightarrow (y, \pperp) \big\}    \Big] \theta[x(1-x)] \theta[y(1-y)],
\end{multline}
where we have defined 
\begin{multline}
D^{-1}(x,\kperp;y,\pperp|M^2) = 
M^2 - \frac{\pperp^2 + m^2}{y} \\ - \frac{(\kperp - \pperp)^2 + \mu^2}{x - y} - \frac{\kperp^2 + m^2}{1-x},
\end{multline}
and taken $\mathbf{R}^\perp = 0$.  Graphically this one-boson exchange potential Eq.~\eqref{OBE} is depicted in Figure \ref{fOBEP}. 

For reference, the bound-state equation \eqref{bse} appears as
\begin{multline} \label{wavefunction}
\psi(x,\kperp) = \dw(x,\kperp|M^2) \int \frac{dy d\pperp}{2(2\pi)^3 y (1-y)} \\ \times V(x,\kperp;y,\pperp|M^2) \psi(y,\pperp)
\end{multline}

\subsection{Form factors}
\begin{figure}
\begin{center}
\epsfig{file=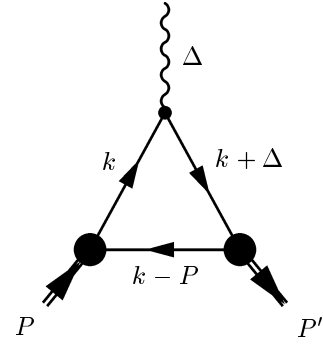}
\caption{The leading-order diagram for the electromagnetic form factor}
\label{ftri}
\end{center}
\end{figure}

To calculate form factors, we use the electromagnetic vertex $\Gamma^\mu$ constructed in \cite{future} up to the first Born approximation (notice that 
the ladder model's gauged interaction $V^\mu = 0$)
\begin{equation}
\Gamma^\mu =  \Big( \overset{\leftrightarrow}\partial{}^\mu + V G \overset{\leftrightarrow}\partial{}^\mu  \Big) d^{-1}_2,
\end{equation}
where $\overset{\leftrightarrow}\partial{}^\mu$ denotes the electromagnetic coupling to scalars.
% and the factor of $2$ originates from identical constituents.

Now using Eq.~\eqref{324} to first order in $G - \Gt$, the matrix element $J^\mu = \langle \Psi_{P^\prime}| \Gamma^\mu(-\Delta) 
| \Psi_P \rangle$  then appears
\begin{multline}
J^\mu  \approx  \; \langle \gamma_{P^\prime} | \; \Big| G(P^\prime) \Big( 1 + V(P^\prime) ( G(P^\prime) - \Gt(P^\prime)) \Big) 
\\ \times \Big( \overset{\leftrightarrow}\partial{}^\mu (-\Delta)  d_{2}^{-1} + V(-\Delta) G(-\Delta) \overset{\leftrightarrow}\partial{}^\mu (-\Delta) d_{2}^{-1} \Big) 
\\ \times \Big( 1 + (G(P) - \Gt(P)) V(P)  \Big) G(P) \Big| \; | \gamma_{P} \rangle  \\
	=  \Big( J^\mu_{\text{LO}} + \delta J^\mu_{i} +  \delta J^\mu_{f} + \delta J^\mu_{\gamma} \Big) + \mathcal{O}[V^2],
\end{multline}
with the leading-order result
\begin{equation} \label{LO}
J^\mu_{\text{LO}} = \langle \gamma_{P^\prime} | \; \Big| G(P^\prime) \overset{\leftrightarrow}\partial{}^\mu (-\Delta)  d_{2}^{-1} G(P) \Big| \; | \gamma_{P} \rangle.
\end{equation}
The first-order terms are
\begin{multline}    
		\delta J^\mu_{i} =  \langle \gamma_{P^\prime} | \; \Big| G(P^\prime) \overset{\leftrightarrow}\partial{}^\mu(-\Delta)  d_{2}^{-1} 
		\\ \times \Big(G(P) - \Gt(P) \Big) V(P) G(P) \Big| \; |\gamma_{P} \rangle \notag
\end{multline}
\begin{multline}
		\delta J^\mu_{f} = \langle \gamma_{P^\prime} | \; \Big| G(P^\prime) V(P^\prime) 
		\\ \times \Big(G(P^\prime) - \Gt(P^\prime) \Big) 
		\overset{\leftrightarrow}\partial{}^\mu (-\Delta) d_{2}^{-1} G(P) \Big| \; | \gamma_{P} \rangle  \notag
\end{multline}
\begin{multline}
		\delta J^\mu_{\gamma}  = \langle \gamma_{P^\prime} | \; \Big| G(P^\prime) \Big(V(-\Delta) G(-\Delta) 
		\overset{\leftrightarrow}\partial{}^\mu (-\Delta) \Big) \\ \times d_{2}^{-1} G(P) \Big| \; | \gamma_{P} \rangle. \label{NLO}
\end{multline}

As outlined in \cite{future}, Eqs.~\eqref{LO} and \eqref{NLO} can be evaluated by residues being careful to remove two-particle reducible
contributions by utilizing \eqref{wavefunction}. 
%and remain at the correct order perturbatively. 
Here we state the results of these calculations
in $(3+1)$ dimensions. We denote $\Delta^\mu$ as the momentum transfer and define $\Delta^+ = - \zeta P^+$ (see Figure \ref{ftri}).
The leading-order result appears
\begin{widetext}
\begin{equation} \label{FFLO}
J^+_{LO} = - 2 i P^+ \int \frac{\theta(x - \zeta) \; dx \; d\kperp}{2(2\pi)^3 x(1-x) x^\prime} (2 x - \zeta)
\psi^*(x^\prime,\mathbf{k}^{\prime\perp}) \psi(x,\kperp),
\end{equation}
where $x^\prime = \frac{x - \zeta}{1-\zeta}$ and $\mathbf{k}^{\prime\perp} = \kperp + (1-x^\prime) \dperp$ denotes the momentum of the final state.
Using $J^\mu = - i (P + P^\prime)^\mu F(t)$, Eq.~\eqref{FFLO} reduces to the Drell-Yan formula \cite{Drell:1969km} for $\zeta = 0$. 

The first of the leading order corrections is the Born term $\delta J^+_\gamma$. For $x>\zeta$ we have
\begin{multline} \label{bt1}
\delta J^+_{\gamma \; \; (x>\zeta)} = \frac{+ 2iP^+}{(16\pi^3)^2} \int \frac{\theta(x - \zeta) dx d\kperp dy d\pperp (2x - \zeta)}{x x^\prime y (1-y) y^\prime} 
\\ \times \psi^*(y^\prime,\mathbf{p}^{\prime\perp}) D(y^\prime,\mathbf{p}^{\prime\perp};x^\prime,\mathbf{k}^{\prime\perp} |M^2) 
\frac{g^2 \theta(y-x)}{y - x} D(y,\pperp;x,\kperp |M^2) \psi(y,\pperp),
\end{multline}
where $y^\prime = \frac{y - \zeta}{1-\zeta}$ and $\ppp = \pperp + (1- y^\prime) \dperp$. This contribution corresponds to 
diagram $A$ in Figure \ref{ftri2}.  On the other hand, for $x<\zeta$ we have
\begin{figure}
\begin{center}
\epsfig{file=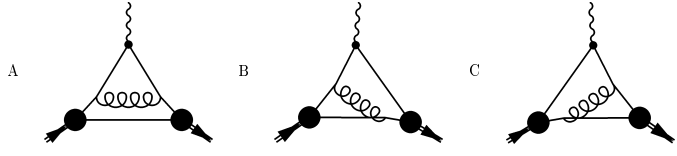,width=3.5in}
\caption{Diagrams which contribute to the form factor at next-to-leading order (for $x >\zeta$).}
\label{ftri2}
\end{center}
\end{figure}
\begin{multline} \label{bt2}
\delta J^+_{\gamma \; \; (x<\zeta)} = \frac{+2 i P^+}{(16\pi^3)^2} \int \frac{\theta(\zeta - x) dx d\kperp dy d\pperp (2 x - \zeta)/\zeta}{y(1-y) \yp \xpp (1-\xpp)} \\ \times
\psi^*(\yp,\ppp) 
\dw(\xpp,\kppp|t) \frac{g^2 \theta(y - x)}{y - x} D(y,\pperp;x,\kperp|M^2) \psi(y,\pperp),
\end{multline}
where $\xpp = x / \zeta$ and $\kppp = \kperp + \xpp \dperp$ denotes the photon's relative momentum. This expression
is diagram $D$ in Figure \ref{fZZZ}.

\begin{figure}
\begin{center}
\epsfig{file=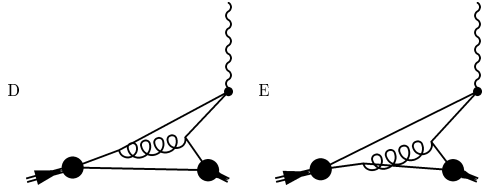,width=2.5in}
\caption{The remaining diagrams (characterized by $x < \zeta$) for the electromagnetic 
		form factor at next-to-leading order.}
\label{fZZZ}
\end{center}
\end{figure}

The next leading-order term is the initial-state iteration $\delta J^+_i$. The only contribution is for $x>\zeta$, namely
\begin{multline} \label{is1}
\delta J^+_{i} = \frac{+2iP^+}{(16\pi^3)^2} \int \frac{\theta(x - \zeta) dx d\kperp dy d\pperp (2x - \zeta)}{x x^\prime (1- x^\prime) y (1-y)}  \\
\times
\psi^*(x^\prime,\mathbf{k}^{\prime\perp}) 
D(y^\prime,\mathbf{p}^{\prime\perp};x^\prime,\mathbf{k}^{\prime\perp} |M^2) 
\frac{g^2 \theta(y-x)}{y - x} D(y,\pperp;x,\kperp |M^2) \psi(y,\pperp),
\end{multline}
which corresponds to diagram $B$ of Figure \ref{ftri2}.

Lastly there is the final-state iteration term $\delta J^+_f$. For $x>\zeta$ we have
\begin{multline}  \label{fs1}
\delta J^+_{f \; \; (x>\zeta)} = \frac{+2iP^+}{(16\pi^3)^2} \int \frac{\theta(x - \zeta) dx d\kperp d\yp d\ppp (2 x - \zeta)}{x(1-x)\xp \yp (1-\yp)}
\\ \times \psi^*(\yp,\ppp) D(\yp,\ppp;\xp,\kpp|M^2) \frac{g^2 \theta(y - x)}{y - x} D(y,\pperp ;x,\kperp |M^2) \psi(x,\kperp), 
\end{multline}
where implicitly $y = \zeta + (1-\zeta) \yp$ and $\pperp = \ppp - (1- \yp) \dperp$. 
This corresponds to diagram $C$ in Figure \ref{ftri}. While for $x<\zeta$, the expression
\begin{multline} \label{fs2}
\delta J^+_{f \; \; (x< \zeta)} = \frac{+2iP^+}{(16\pi^3)^2} \int \frac{\theta(\zeta - x) dx d\kperp d\yp d\ppp (2 x - \zeta)/\zeta}{(1-x) \xpp (1-\xpp) \yp (1-\yp)}
\\ \times \psi^*(\yp,\ppp) \dw(\xpp,\kppp|t) \frac{g^2 \theta(y - x)}{y - x} D(y,\pperp;x,\kperp|M^2) \psi(x,\kperp),
\end{multline}
corresponds to diagram $E$ of Figure \ref{fZZZ}. Eqs.~(\ref{FFLO}-\ref{fs2}) are then the complete 
expressions for the form factor up to first order.
% which we needed above.
\end{widetext}

\section{Old-fashioned time-ordered perturbation theory} \label{oftopt}
The results of this paper can similarly be achieved directly from ``old-fashioned'' time-ordered perturbation theory in a form which utilizes
projecting onto the two-body subspace of the full Fock space. 
%As the light-front Hamiltonian generates light-cone time translation, 
%not surprisingly time-ordered perturbation theory stems from the Hamiltonian.
For a nice, complete discussion of this formalism for the light-cone ladder model, see \cite{Cooke:2000ef}. 
In this Appendix, we show how to derive higher Fock space components in this formalism, 
thereby demonstrating the generation of higher components from the lowest sector we found indirectly for GPDs 
and form factors in section \ref{gpds}. 

We write the light-cone Hamiltonian as a sum of a free piece and an interacting piece which carries an explicit power of the weak coupling $g$. 
In an obvious notation this is
\begin{equation}
P^- = P^-_o + g P_{I}^-.
\end{equation}
The free term $P^-_o$ is diagonal in the Fock state basis, while the interaction generally mixes components of different particle number (in the scalar model we consider above, the interaction is completely off-diagonal since there are no instantaneous terms). Let us suppose that in the full Fock basis, we have an eigenstate of the Hamiltonian, i.e.
\begin{equation}
\Big( P^-_o + g P_{I}^- \Big) \psiket = \pminus \psiket, 
\end{equation}
where the eigenvalue is labeled by $\pminus$. Since the coupling is presumed small, the mixing of Fock components with a large number of 
particles will be small. Thus one imagines our bound state will be dominated by the two-body Fock component. 

To make this observation formal, we define projections operators on the Fock space $\pe$ and $\qu$ in the usual sense. 
The operator $\pe$ projects out only the two-particle subspace of the full Fock space and hence $\qu$ projects out the compliment. Let us define the action
of these operators on our eigenstate
\begin{align}
\pe \psiket & = \psitwoket \label{2ket}\\
\qu \psiket & = \psiquket  \label{qket}.
\end{align}
%Given $\pe + \qu = 1$, we have immediately that $\psiket = \psitwoket + \psiquket$. We now wish to derive an equation for the two-body state
%in terms of effective two-body operators. First we note
%\begin{equation} \label{start}
%\pe P^- \psiket = \pminus \pe \psiket = \pminus \psitwoket,
%\end{equation}
%and then write $\pe P^- 1 = \pe P^- (\pe + \qu) = P^-_{\pe \pe} \pe + P^-_{\pe \qu} \qu$, where we have defined the following notation
%for any operators $A$ and $B$, $P^-_{A B} \equiv A P^- B$. Using this operator relation in Eq. \eqref{start} yields
%\begin{equation} \label{one}
%P^-_{\pe \qu} \psiquket = \Big( \pminus - P^-_{\pe \pe}  \Big) \psitwoket.
%\end{equation} 
%Now we use the same procedure on the state $\qu P^- \psiket$ to find
%\begin{equation} \label{two}
%P^-_{\qu \pe} \psitwoket = \Big( \pminus - P^-_{\qu \qu}  \Big) \psiquket.
%\end{equation}

%Combining Eqs. \eqref{one} and \eqref{two}, we arrive at the following equation for the two-body Fock component
As is well known, combination of Eqs.~\eqref{2ket} and \eqref{qket} leads to the following equation for the two-body Fock component
\begin{equation}
P^-_{\text{eff}} \psitwoket = \pminus \psitwoket,
\end{equation}
where the effective two-body Hamiltonian is
\begin{align} \label{veff}
P^-_{\text{eff}} & \equiv P^-_{\pe \pe} + V_{\text{eff}} \\
	& = P^-_{\pe \pe} + P^-_{\pe \qu} \frac{1}{\pminus - P^-_{\qu \qu}} P^-_{\qu \pe},
\end{align}
and we have defined the following notation for any operators $A$ and $B$, $P^-_{A B} \equiv A P^- B$.
The effective two-body interaction $V_{\text{eff}}$ defined in equation \eqref{veff} is dependent upon the energy eigenvalue $\pminus$ since we have 
suppressed the degrees of freedom of the $\qu$ subspace. The relation between the $\qu$-space probability (i.e. the non-valence contribution)
and the effective interaction appears
\begin{equation} \label{quspace}
\langle \psi_{\qu} \psiquket = - \frac{\partial}{\partial \pminus} \psitwobra V_{\text{eff}} \psitwoket.
\end{equation}

In a weak-binding limit, we can series expand the effective interaction in powers of the coupling and thereby re-derive the light-front potential. 
Given that every boson emitted must be absorbed in the two-quark sector, we can have only an even number of interactions and hence
\begin{equation}
V_{\text{eff}} = \pe g P^-_{I} \qu \frac{1}{\pminus - P_{o}^-} \sum_{n=0}^{\infty} \Bigg( \frac{g P^-_{I}}{\pminus - P^-_{o}} \Bigg)^{2 n} \qu g P^-_{I} \pe.
\end{equation}
So, for example, at leading order we have all possible ways to propagate from the two-body sector and back with only two interactions in between. The diagrams in Figure \ref{fOBEP} correspond to the two possibilities distinguished by the action of $\frac{1}{\pminus - P^-_{o}}$ between interactions. At the next order, we have all possible ways to propagate from two bodies to two bodies with four interactions in between, \emph{etc}.

 To generate higher Fock components from the two-body sector, we necessarily must look at the $\qu$-space state
arrived at from Eqs.~\eqref{2ket} and \eqref{qket}
 \begin{equation}
 \psiquket = \frac{1}{\pminus - P^-_{\qu \qu}} P^-_{\qu \pe} \psitwoket. 
 \end{equation}   
 To generate an $n$-body Fock component from this state, we merely act with an $n$-body projection operator which we shall denote $\qu_{n}$. 
Similar to the above, we expand in powers of the coupling to find
\begin{equation}
| \psi_{n} \rangle = \qu_{n} \frac{1}{\pminus - P^-_{o}} \sum_{n = 0}^{\infty} \Bigg(  \frac{g P^-_{I}}{\pminus - P^-_{o}} \Bigg)^n \qu g P^-_{I} \psitwoket.
\end{equation}
For example, the leading-order three-body state is obtained by attaching a boson to a quark line in the only two possible ways 
(and adding the light-front energy denominator at the end). With these three-body states, we can consider all possible three-to-three overlaps that 
would contribute to the form factor. These are depicted in Figure \ref{ftri2} (with the exception of a quark 
self-interaction).  The four-body sector is richer since there are two-boson, two-quark states as well as four-quark states. 
The two-to-four overlaps required for GPDs must have four quarks. At leading order, we generate the diagrams encountered above in 
Figure \ref{fZZZ}. Not surprisingly directly applying time-ordered perturbation theory from a light-front Hamiltonian agrees with our 
derivation above from covariant perturbation theory in the Bethe-Salpeter formalism.


\begin{thebibliography}{99}

%\cite{Dirac:1949cp}
\bibitem{Dirac:1949cp}
P.~A.~M.~Dirac,
%``Forms Of Relativistic Dynamics,''
Rev.\ Mod.\ Phys.\ {\bf 21}, 392 (1949).
%%CITATION = RMPHA,21,392;%%

%\cite{Leutwyler:1977vy}
\bibitem{Leutwyler:1977vy}
H.~Leutwyler and J.~Stern,
%``Relativistic Dynamics On A Null Plane,''
Annals Phys.\  {\bf 112}, 94 (1978).
%%CITATION = APNYA,112,94;%%

%\cite{Lepage:1980fj}
\bibitem{Lepage:1980fj}
G.~P.~Lepage and S.~J.~Brodsky,
%``Exclusive Processes In Perturbative Quantum Chromodynamics,''
Phys.\ Rev.\ D {\bf 22}, 2157 (1980).
%%CITATION = PHRVA,D22,2157;%%

%\cite{Brodsky:1997de}
\bibitem{Brodsky:1997de}
S.~J.~Brodsky, H.~C.~Pauli and S.~S.~Pinsky,
%``Quantum chromodynamics and other field theories on the light cone,''
Phys.\ Rept.\  {\bf 301}, 299 (1998).
%[arXiv:hep-ph/9705477].
%%CITATION = HEP-PH 9705477;%%

%\cite{Carbonell:1998rj}
\bibitem{Carbonell:1998rj}
J.~Carbonell, B.~Desplanques, V.~A.~Karmanov and J.~F.~Mathiot,
%``Explicitly covariant light-front dynamics and relativistic few-body  systems,''
Phys.\ Rept.\  {\bf 300}, 215 (1998).
%[arXiv:nucl-th/9804029].
%%CITATION = NUCL-TH 9804029;%%

%\cite{Miller:2000kv}
\bibitem{Miller:2000kv}
G.~A.~Miller,
%``Light front quantization: A technique for relativistic and realistic  nuclear physics,''
Prog.\ Part.\ Nucl.\ Phys.\  {\bf 45}, 83 (2000).
%[arXiv:nucl-th/0002059].
%%CITATION = NUCL-TH 0002059;%%

\bibitem{future}
B.~C.~Tiburzi and G.~A.~Miller, hep-ph/0210304.

%\cite{Tiburzi:2002mn}
\bibitem{Tiburzi:2002mn}
B.~C.~Tiburzi and G.~A.~Miller,
%``Light front Bethe-Salpeter equation applied to form factors,  generalized parton distributions and generalized distribution  amplitudes,''
hep-ph/0205109.
%%CITATION = HEP-PH 0205109;%%

%\cite{Bakker:2000rd}
\bibitem{Bakker:2000rd}
B.~L.~Bakker and C.-R.~Ji,
%``Disentangling intertwined embedded-states and spin effects in  light-front quantization,''
Phys.\ Rev.\ D {\bf 62}, 074014 (2000).
%[arXiv:hep-th/0003105].
%%CITATION = HEP-TH 0003105;%%

%\cite{Sales:1999ec}
\bibitem{Sales:1999ec}
J.~H.~Sales, T.~Frederico, B.~V.~Carlson and P.~U.~Sauer,
%``Light-front Bethe-Salpeter equation,''
Phys.\ Rev.\ C {\bf 61}, 044003 (2000);
%[arXiv:nucl-th/9909029].
%%CITATION = NUCL-TH 9909029;%%
%\cite{Sales:2001gk}
%\bibitem{Sales:2001gk}
%J.~H.~Sales, T.~Frederico, B.~V.~Carlson and P.~U.~Sauer,
%``Renormalization of the ladder light-front Bethe-Salpeter equation in the Yukawa model,''
%Phys.\ Rev.\ C {\bf 63}, 
064003 (2001).
%%CITATION = PHRVA,C63,064003;%%

%\cite{Itzykson:rh}
\bibitem{Itzykson:rh}
C.~Itzykson and J.~B.~Zuber,
{\it Quantum Field Theory},
New York, USA: McGraw-Hill (1980) 705 p.(International Series In Pure and Applied Physics).

%\cite{Woloshyn:wm}
\bibitem{Woloshyn:wm}
R.~M.~Woloshyn and A.~D.~Jackson,
%``Comparison Of Three-Dimensional Relativistic Scattering Equations,''
Nucl.\ Phys.\ B {\bf 64}, 269 (1973).
%%CITATION = NUPHA,B64,269;%%

%\cite{Muller:1994fv}
\bibitem{Muller:1994fv}
D.~M\"uller, D.~Robaschik, B.~Geyer, F.~M.~Dittes and J.~Ho\v{r}ej\v{s}i,
%``Wave functions, evolution equations and evolution kernels from light-ray operators of {QCD},''
Fortsch.\ Phys.\ {\bf 42}, 101 (1994)
[hep-ph/9812448];
%%CITATION = HEP-PH 9812448;%%
%\cite{Ji:1997ek}
%\bibitem{Ji:1997ek}
X.-D.~Ji,
%``Gauge invariant decomposition of nucleon spin,''
Phys.\ Rev.\ Lett.\ {\bf 78}, 610 (1997);
%[hep-ph/9603249].
%%CITATION = HEP-PH 9603249;%%
%\cite{Ji:1997nm}
%\bibitem{Ji:1997nm}
X.-D.~Ji,
%``Deeply-virtual Compton scattering,''
Phys.\ Rev.\ D {\bf 55}, 7114 (1997);
%[hep-ph/9609381].
%%CITATION = HEP-PH 9609381;%%
%\cite{Radyushkin:1997ki}
%\bibitem{Radyushkin:1997ki}
A.~V.~Radyushkin,
%``Nonforward parton distributions,''
Phys.\ Rev.\ D {\bf 56}, 5524 (1997);
%[hep-ph/9704207].
%%CITATION = HEP-PH 9704207;%%
%\cite{Radyushkin:1996ru}
%\bibitem{Radyushkin:1996ru}
A.~V.~Radyushkin,
%``Asymmetric gluon distributions and hard diffractive electroproduction,''
Phys.\ Lett.\ B {\bf 385}, 333 (1996);
%[hep-ph/9605431].
%%CITATION = HEP-PH 9605431;%%
%\cite{Collins:1997fb}
%\bibitem{Collins:1997fb}
J.~C.~Collins, L.~Frankfurt and M.~Strikman,
%``Factorization for hard exclusive electroproduction of mesons in QCD,''
Phys.\ Rev.\ D {\bf 56}, 2982 (1997);
%[hep-ph/9611433].
%%CITATION = HEP-PH 9611433;%%
%\cite{Radyushkin:1998rt}
%\bibitem{Radyushkin:1998rt}
A.~V.~Radyushkin,
%``Nonforward parton densities and soft mechanism for form factors and  wide-angle Compton scattering in {QCD},''
Phys.\ Rev.\ D {\bf 58}, 114008 (1998).
%[arXiv:hep-ph/9803316].
%%CITATION = HEP-PH 9803316;%%

%\cite{Brodsky:2001xy}
\bibitem{Brodsky:2001xy}
S.~J.~Brodsky, M.~Diehl and D.~S.~Hwang,
%``Light-cone wavefunction representation of deeply virtual Compton  scattering,''
Nucl.\ Phys.\ B {\bf 596}, 99 (2001);
%[hep-ph/0009254].
%%CITATION = HEP-PH 0009254;%%
%\cite{Diehl:2001xz}
%\bibitem{Diehl:2001xz}
M.~Diehl, T.~Feldmann, R.~Jakob and P.~Kroll,
%``The overlap representation of skewed quark and gluon distributions,''
Nucl.\ Phys.\ B {\bf 596}, 33 (2001).
%[hep-ph/0009255].
%%CITATION = HEP-PH 0009255;%%

%\cite{Diehl:1997bu}
\bibitem{Diehl:1997bu}
M.~Diehl, T.~Gousset, B.~Pire and J.~P.~Ralston,
%``Testing the handbag contribution to exclusive virtual Compton  scattering,''
Phys.\ Lett.\ B {\bf 411}, 193 (1997).
%[arXiv:hep-ph/9706344].
%%CITATION = HEP-PH 9706344;%%

%\cite{Burkardt:2002uc}
\bibitem{Burkardt:2002uc}
M.~Burkardt, X.-D.~Ji and F.~Yuan,
%``Scale dependence of hadronic wave functions and parton densities,''
hep-ph/0205272.
%%CITATION = HEP-PH 0205272;%%

%\cite{Tiburzi:2002kr}
\bibitem{Tiburzi:2002kr}
B.~C.~Tiburzi and G.~A.~Miller,
%``Generalized parton distributions for q anti-q pions,''
hep-ph/0209178.
%%CITATION = HEP-PH 0209178;%%

%\cite{Antonuccio:1997tw}
\bibitem{Antonuccio:1997tw}
F.~Antonuccio, S.~J.~Brodsky and S.~Dalley,
%``Light-cone wavefunctions at small x,''
Phys.\ Lett.\ B {\bf 412}, 104 (1997).
%[arXiv:hep-ph/9705413].
%%CITATION = HEP-PH 9705413;%%

\bibitem{Diehl:1998dk}
M.~Diehl, T.~Gousset, B.~Pire and O.~Teryaev,
%``Probing partonic structure in gamma* gamma --> pi pi near threshold,''
Phys.\ Rev.\ Lett.\ {\bf 81}, 1782 (1998);
%[hep-ph/9805380]
%%CITATION = HEP-PH 9805380;%%
%\cite{Diehl:2000uv}
%\bibitem{Diehl:2000uv}
M.~Diehl, T.~Gousset and B.~Pire,
%``Exclusive production of pion pairs in gamma* gamma collisions at large  Q**2,''
Phys.\ Rev.\ D {\bf 62}, 073014 (2000);
%[hep-ph/0003233].
%%CITATION = HEP-PH 0003233;%%
%\cite{Polyakov:1999ze}
M.~V.~Polyakov,
%``Study of two-pion light-cone distribution amplitudes in the resonance  region and at low energies,''
Nucl.\ Phys.\ B {\bf 555}, 231 (1999);
%%CITATION = HEP-PH 9809483;%%
%\cite{Lehmann-Dronke:1999aq}
%\bibitem{Lehmann-Dronke:1999aq}
B.~Lehmann-Dronke, P.~V.~Pobylitsa, M.~V.~Polyakov, A.~Sch\"afer and K.~Goeke,
%``Hard diffractive electroproduction of two pions,''
Phys.\ Lett.\ B {\bf 475}, 147 (2000),
%[arXiv:hep-ph/9910310].
%\cite{Diehl:2001fv}
%\bibitem{Diehl:2001fv}
M.~Diehl, P.~Kroll and C.~Vogt,
%``The handbag contribution to gamma gamma $\to$ pi pi and K K,''
Phys.\ Lett.\ B {\bf 532}, 99 (2002).
%[arXiv:hep-ph/0112274].
%%CITATION = HEP-PH 0112274;%%

%\cite{Cooke:2000ef}
\bibitem{Cooke:2000ef}
J.~R.~Cooke and G.~A.~Miller,
%``Ground states of the Wick-Cutkosky model using light-front dynamics,''
Phys.\ Rev.\ C {\bf 62}, 054008 (2000).
%[arXiv:nucl-th/0002016].
%%CITATION = NUCL-TH 0002016;%%

%\cite{Drell:1969km}
\bibitem{Drell:1969km}
S.~D.~Drell and T.-M.~Yan,
%``Connection Of Elastic Electromagnetic Nucleon Form-Factors At Large Q**2 And Deep Inelastic Structure Functions Near Threshold,''
Phys.\ Rev.\ Lett.\  {\bf 24}, 181 (1970);
%%CITATION = PRLTA,24,181;%%
%\cite{West:1970av}
%\bibitem{West:1970av}
G.~B.~West,
%``Phenomenological Model For The Electromagnetic Structure Of The Proton,''
Phys.\ Rev.\ Lett.\  {\bf 24}, 1206 (1970).
%%CITATION = PRLTA,24,1206;%%
\end{thebibliography}
\end{document}